\documentclass[11pt,twocolumn]{article} % Specifies the document style.
\usepackage{graphicx}
\usepackage{amssymb}
\usepackage{authblk}
\usepackage{abstract}
\RequirePackage{amsmath}
\RequirePackage{fix-cm}
\RequirePackage{graphicx}
\RequirePackage{latexsym}
\RequirePackage{xspace}

\newcommand{\Fermi}{{\sl Fermi}\xspace}
\newcommand{\FermiLAT}{{\sl Fermi}/LAT\xspace}

\newcommand{\SwiftUVOT}{{\sl Swift}/UVOT\xspace}

\newcommand{\IceCube}{{\sc IceCube}\xspace}
\newcommand{\ANTARES}{\textsc{Antares}\xspace}
\newcommand{\Antares}{\textsc{Antares}\xspace}
\newcommand{\tauz}{\ensuremath{\tau_{\rm z}}\xspace}
\newcommand{\tauliv}{\ensuremath{\tau_{\rm LIV}}\xspace}
\newcommand{\LIV}{{\rm LIV}}
\newcommand{\Psiz}{\ensuremath{\psi_{\rm z}}\xspace}
\newcommand{\Psiliv}{\ensuremath{\psi_{\rm LIV}}\xspace}
\newcommand{\psiliv}{\Psiliv}
\newcommand{\psiz}{\Psiz}
\newcommand{\psilivmeas}{\ensuremath{\psi_{\rm LIV, meas}}\xspace}
\newcommand{\psizmeas}{\ensuremath{\psi_{\rm z, meas}}\xspace}
\newcommand{\psimeas}{\ensuremath{\psi_{\rm meas}}\xspace}
\newcommand{\deltamax}{\ensuremath{\delta_{\rm max}}\xspace}
\newcommand{\taumax}{\ensuremath{\tau_{\rm max}}\xspace}
\newcommand{\MDP}{\ensuremath{\mathcal{MDP}}\xspace}
\newcommand{\unit}[1]{\ensuremath{\, \mathrm{#1}\xspace}}

\newcommand{\s}{\unit{s}}

\newcommand{\GeV}{\unit{GeV}}

\newcommand{\e}[1]{\ensuremath{\cdot 10^{#1}}}

\renewcommand{\deg}{\ensuremath{^{\circ}}\xspace}

\date{\today}
\author[1]{S.~Adri\'an-Mart\'inez  }
\author[2]{A.~Albert }
\author[4]{M.~Andr\'e  }
\author[5]{M.~Anghinolfi}
\author[6]{G.~Anton  }
\author[1]{M.~Ardid  }
\author[3]{J.-J.~Aubert }
\author[9]{B.~Baret \footnote{\scriptsize{Corresponding authors. Email addresses: j.schmid.phys@gmail.com (J.~Schmid) \& baret@in2p3.fr (B.~Baret)}}}
\author[10]{J.~Barrios-Marti  }
\author[5]{S.~Basa }
% ...

\affil[1]{\scriptsize{Institut d'Investigaci\'o per a la Gesti\'o Integrada de les Zones Costaneres (IGIC) - Universitat Polit\`ecnica de Val\`encia. C/  Paranimf 1, 46730 Gandia, Spain.} }
\affil[2]{\scriptsize{GRPHE - Institut universitaire de technologie de Colmar, 34 rue du Grillenbreit BP 50568 - 68008 Colmar, France } }
\affil[3]{\scriptsize{CPPM, Aix-Marseille Universit\'e, CNRS/IN2P3, Marseille, France} }
\affil[4]{\scriptsize{Technical University of Catalonia, Laboratory of Applied Bioacoustics, Rambla Exposici\'o,08800 Vilanova i la Geltr\'u,Barcelona, Spain} }
\affil[5]{\scriptsize{INFN - Sezione di Genova, Via Dodecaneso 33, 16146 Genova, Italy} }
\affil[6]{\scriptsize{Friedrich-Alexander-Universit\"at Erlangen-N\"urnberg, Erlangen Centre for Astroparticle Physics, Erwin-Rommel-Str. 1, 91058 Erlangen, Germany} }
% ...
\affil[9]{\scriptsize{APC, Universit\'e Paris Diderot, CNRS/IN2P3, CEA/IRFU, Observatoire de Paris, Sorbonne Paris Cit\'e, 75205 Paris, France} }
\affil[10]{\scriptsize{IFIC - Instituto de F\'isica Corpuscular, Edificios Investigaci\'on de Paterna, CSIC - Universitat de Val\`encia, c/ Catedr\'atico Jos\'e Beltr\'an, 2, Paterna 46980, Valencia, Spain} }
% ...

\title{Stacked search for time shifted high energy neutrinos from gamma ray bursts with the \ANTARES neutrino telescope.%\thanksref{t1}
}

\begin{document}
\onecolumn
\maketitle
\clearpage
\twocolumn[
%\begin{@twocolumnfalse}

%\newpage
%\begin{abstract}
\begin{onecolabstract}
A search for high-energy neutrino emission correlated with gamma-ray bursts outside the electromagnetic prompt-emission time window is presented.
Using a stacking approach of the time delays between reported gamma-ray burst alerts and spatially coincident muon-neutrino signatures, data from the \ANTARES neutrino telescope recorded between 2007 and 2012 are analysed.  One year of public data from the \IceCube detector between 2008 and 2009 have been also investigated. The respective timing profiles are scanned for statistically significant accumulations within 40 days of the Gamma Ray Burst, as expected from Lorentz Invariance Violation effects and  some astrophysical models.
No significant excess over the expected accidental coincidence rate could be found in either of the two data sets. %with marginally significant evidence in the one-year IceCube data sample at the 1.9$\sigma$ level.
The average strength of the neutrino signal is found to be fainter than one detectable neutrino signal per hundred gamma-ray bursts in the \ANTARES data at 90\% confidence level.

%\keywords{neutrinos - gamma-ray burst: general - multimessenger astronomy - Lorentz Invariance Violation}

\end{onecolabstract}
%\end{abstract}
%\nopagebreak
%\end{@twocolumnfalse}
] %\twocolumn[
%\saythanks
%\cleardoublepage
%\newpage
%\newpage
%\twocolumn
\section{Introduction}
Gamma-ray bursts (GRBs) are among the most powerful sources in the universe, which makes them suitable candidates for the acceleration of the highest-energy cosmic rays. 
Unambiguous evidence for the acceleration of hadrons in astrophysical environments can be provided by the detection of neutrinos that would be coincidently produced when accelerated protons interact with the ambient photon field (see, e.g. \cite{Waxman95a},\cite{Waxman97a},\cite{Waxman00a} and references therein). 
Searches for the  emission of neutrinos from GRBs have been performed by a variety of experiments, for instance  Super-Kamiokande \cite{Super_Kamiokande00a}, AMANDA \cite{Amanda08a},  Baikal \cite{Baikal11a}, RICE \cite{Rice07a}, ANITA \cite{Anita11a}, IceCube \cite{IceCube10a,IceCube12a} and\Antares \cite{Antares13a,Antares13b}. While covering a wide range of neutrino energies these studies have so far focussed mainly on the time window coincident with the electromagnetic signal of GRBs.  
 Up to now no neutrino signal could be identified by any neutrino detector during the prompt emission phases, and %the first optimistic
 analytical models from \cite{Guetta04a} based on \cite{Waxman97a} have already been excluded by the IceCube collaboration \cite{IceCube12a}. 
There has also been some effort to successively extend the search time windows in the IceCube data from [-1h,+3h] up to $\pm 1$ day \cite{IceCube10a,IceCube12a}, and up to $\pm 15$ days  \cite{Casey13a}, to account for prolonged neutrino emission.
%Three neutrino candidates had been found with arrival directions roughly correlated with reported GRBs but in the order of $\mathcal{O}(10^3)$~s before the respective GRB  in the IceCube data \cite{IceCube12a, Whitehorn12a}.
%In \cite{Casey13a}, the most significant neutrino coincidence was found to be $4.9$ days before a GRB alert announced by Fermi:GBM. 
However none of these searches could bring compelling evidence for a GRB signal, since all detected events have been identified with cosmic-ray induced air showers or were of low significance because of the large time windows. \\
While the search for a signal of neutrinos coincident with the emission of high-energy photons is the most common ansatz, there are many models that predict time-shifted neutrino signals, 
such as neutrino precursors \cite{Razzaque03a}, afterglows (e.g. \cite{Waxman00b}), or different Lorentz Invariance Violation (LIV) effects for photons and neutrinos on their way to Earth \cite{Amelino09a,Jacob07a}. 
For instance, the possibility that three low significance neutrino-like events found in the IceCube data \cite{IceCube12a} could have been produced by GRBs but arrived before the photon signal due to LIV effects is discussed in \cite{Amelino15a}. 
%actual gamma-ray-burst neutrinos with time-shifted arrival times due to Lorentz Invariance Violation effect.
Probing such scenarios requires a new approach to the search for correlated emission. %, since the most simple generalisation of the method, an extended symmetric time window, accumulates background linearly to the size of the window and thus prevents a weak signal to protrude significantly. 
Moreover, in all aforementioned scenarios, the neutrino signal is simply shifted in time with respect to the electromagnetic signal, and none of these models predict any considerably prolonged neutrino emission. 
%It is likely that a presumably detectable signal must be weak (between zero and one neutrino per GRB), as otherwise it would have been discovered in previous searches for neutrino point sources or event multiplets. 
Hence the approach used in this paper and described in section \ref{sec:principle} aims at identifying a presumably faint neutrino signal that is shifted %\footnote{Here and in the following, the term 'delay' will denote both negative and positive shifts in time of the neutrino emission with respect to the electromagnetic signal.}
 with respect to the electromagnetic GRB emission by an unknown time offset, while making no assumption about the origin of such an offset. 
%%%%%%

%%%%%%
% PROPOSITION
%%%%%%

\section{Neutrino candidates and GRB sample}
%\subsection{The\Antares neutrino telescope}
Neutrino telescopes are arrays of photomultipliers deployed in a very large volume of transparent medium like Antarctic ice or deep-sea water. They detect the Cherenkov light generated by the products of the interaction of a high-energy neutrino in the vicinity of the detector.
The direction of the impinging neutrino is reconstructed using the timing of signals from photomultipliers, while the detected amount of light gives an estimate of the neutrino energy.
The\Antares telescope \cite{Antares11a} is located at a depth of $2475~\mathrm{m}$ in the French Mediterranean Sea off the coast of Toulon, at $42^{\circ}48'~$N, $6^{\circ}10'~$E. It comprises 885 optical modules housing 10'' photomultipliers in 17'' glass spheres installed on 12 strings, representing an instrumented volume of $0.02~\mathrm{km^3}$.\\
The following analysis focuses on the detection of muon trajectories from below the horizon, which are produced by muon-neutrino charged-current interactions. This channel provides significantly better directional reconstruction than neutral-current interactions and charged current interactions from the other neutrino flavors. %In addition, the number of photon hits in the optical modules that have been used to reconstruct the particle track direction can serve as a simple energy estimator of a neutrino event.
In this channel,\Antares is the most sensitive detector for sources in a large part of the southern sky up to a few 100~TeV \cite{Antares15}. 
%In this analysis, the number of detected photons in the optical modules that have been used to reconstruct the particle track direction will be used as a simple energy estimator of a presumable neutrino event. 
 
\begin{figure} \centering
	\includegraphics[width=0.494\textwidth]{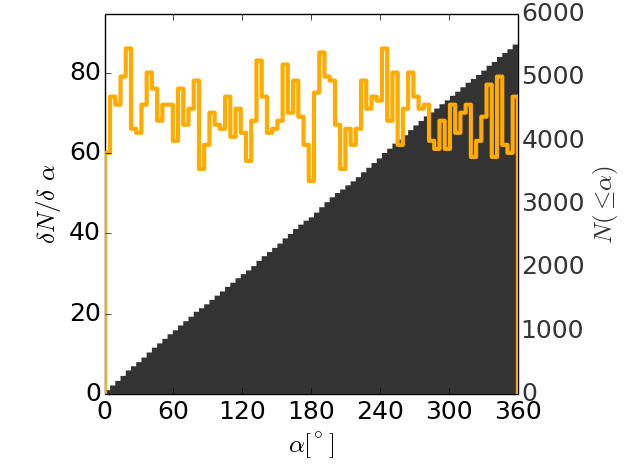}
	\caption{\label{fig:event_distributions} %Distributions of time in Modified Julian Date ({\sl upper panel}) and right ascension $\alpha$ ({\sl lower panel}) of the\Antares neutrino event sample (March 2007 -- 2012). The respective cumulative distributions are shown in black.}
	Distribution of the right ascension $\alpha$ of the\Antares neutrino event sample (March 2007 -- December 2012). The respective cumulative distribution is shown in black.}
\end{figure}

\begin{figure} \centering 
\includegraphics[width=0.48\textwidth]{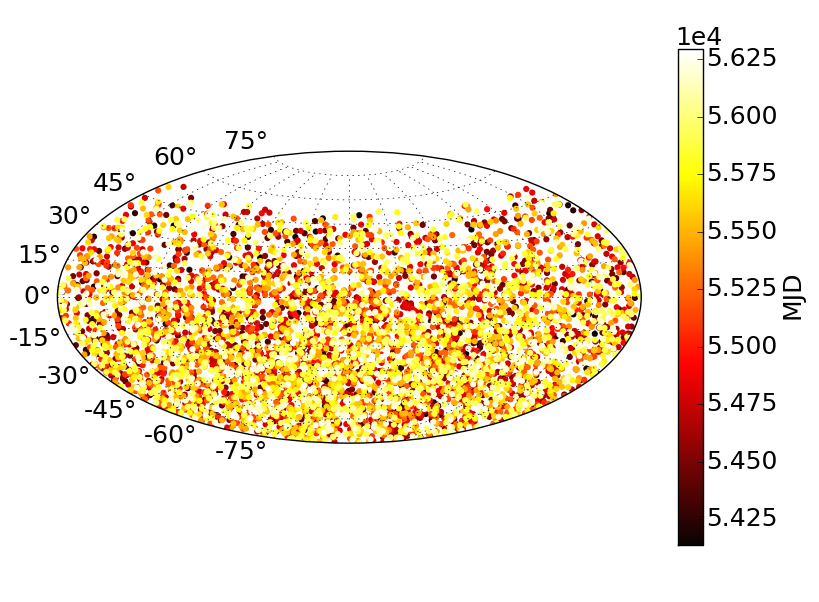}
\includegraphics[width=0.48\textwidth]{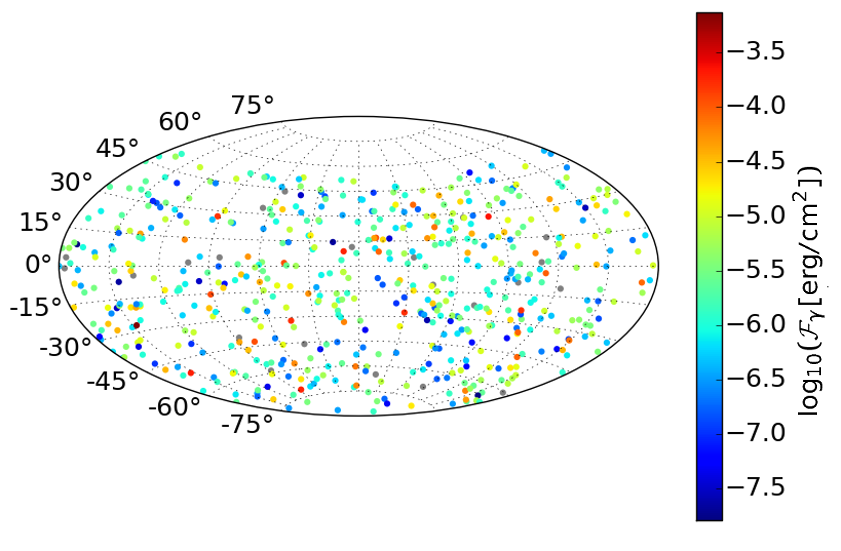}
\caption{\label{fig:skymaps}Distributions in equatorial coordinates of selected GRBs  {\sl (upper panel)} and recorded neutrino candidates {\sl (lower panel)} for the\ANTARES event sample. Each GRB's location is color-coded with the photon fluence $\mathcal{F}_\gamma$; those with no measurement are coloured in gray. 
The color of neutrino events represents their detection time.}
\end{figure}

%
%\subsection{Neutrino event sample}
%If one wants to remain sensitive to the widest possible range of phenomena, the usual methods of selecting high-energy neutrino candidates by optimizing event selection criteria for the best limit or detection probability given a particular signal model are not applicable. 
%Samples of presumable neutrino events that have been singled out in searches for neutrino point sources provide naturally suited data for this kind of approach, as the stringent selection quality criteria ensure efficient suppression of the background contamination from falsely reconstructed muons produced in cosmic-ray air showers, and yield at the same time excellent angular resolution as required for the search for directional coincidences. 
The sample of\Antares events used in this analysis consists of 5516 neutrino candidates selected from data collected between March 2007 and the end of 2012 \cite{Antares14c}. 
%The stringent selection quality criteria ensure efficient suppression of the background contamination from falsely reconstructed muons produced in cosmic-ray air showers
%atmospheric muons. 
%In addition, they ensure an excellent angular resolution suited to the search for directional coincidences.  
%The sample used in this analysis consist of 5516 neutrino candidates selected from the connection of the first detection lines in March 2007 to the end of 2012 \cite{Antares14c}. 
From Monte Carlo simulations the angular resolution, defined as the median of the space angle $\delta_{\rm err}$ between the true and reconstructed direction of neutrinos for an $E^{-2}$ differential spectrum, is $0.38\deg$, with a contamination from atmospheric muons of $10\%$. 
%The respective time and right-ascension distributions of the neutrino candidates are shown in Figure~\ref{fig:event_distributions}. 
The right-ascension distribution of the neutrino candidates is shown in Figure~\ref{fig:event_distributions}. \\
%
%
%\subsection{GRB sample}
A suitable GRB sample was consolidated similarly to the one used in \cite{Antares13b}. 
It was built using catalogs from the \textsl{Swift} \cite{Gehrels04a} and \textsl{Fermi} satellites \cite{Atwood09a,Meegan09a}, and  supplemented by a table from the \IceCube Collaboration\footnote{available on-line at \url{http://grbweb.icecube.wisc.edu/} } \cite{IceCube11a}, with information parsed from the GRB Coordinates Network (GCN) notices. 
Since only the time and position information (and the measured redshift, if available) of each announced GRB was used, no further selection on e.g. the quality of the spectral measurements was required, leading to 1488 GRBs.
Only GRB alerts were taken into account that occurred both below the horizon of the neutrino telescope and during the covered neutrino data collection period.
%However, to avoid any boundary effects, GRBs at the beginning and the end of the livetime of the neutrino data were excluded. 
% to ensure symmetrical search time windows.
%
The upper panel of Figure~\ref{fig:skymaps} shows the distribution of the selected neutrino candidates, which are homogeneously distributed in time. The lower panel displays the accordingly selected GRBs in equatorial coordinates and their measured fluence. 
%
%%%%%%%%
\section{Principle of the search}
\label{sec:principle} 
Neutrino signatures are searched-for in an angular cone around the direction of, and  within a maximum time offset from the time of each GRB.
%For each GRB with given time and position in the sky, the presence of a neutrino within a certain angular cone and  within a maximum time offset is searched for.
For any such space and time coincidence, the time difference with respect to the GRB alert is recorded. In order to avoid any boundary effect such as an artificial asymetry of neutrino candidates around a GRB alert close to the beginning or end of neutrino telescope data taking, GRBs detected during a period of half the considered maximum time offset at the beginning and end of the neutrino data sample are excluded. 
%For each GRB, the presence of a neutrino in the candidate list within a maximum time interval and certain angular cone is searched for. For any such  coincidence, the time difference with respect to the gamma-ray-burst alert is recorded.
The collected time differences are stacked in a common timing profile.
In the case of no signal, only purely accidental spatial coincidences of the neutrino candidates with the defined search cones around the GRB positions would be expected. 
The observed time shifts should then be distributed randomly, yielding a flat stacked  distribution where all shifts are equally likely. 
Any neutrino emission associated with the GRBs, even if faint, can give rise to a cumulative effect in these stacked profiles, which can then be identified by its discrepancy from the background hypothesis. 
An optimal choice of the search cone size \deltamax naturally depends on the GRB's position accuracy and the neutrino direction reconstruction uncertainty. The size of the probed time window \taumax should be defined as the largest shift predicted by any of the models under consideration. 
	Such a procedure had already been proposed \cite{Eijndhoven08a}, where windows of $\pm 1\unit{h}$ around the GRB satellite triggers under study were considered.
	The approach presented in the following is extended to allow significantly larger time windows and different origins of the time shift.
%
%%%%%%%
This method is intrinsically different from those previously developed by the \IceCube ~Collaboration \cite{IceCube12a}\cite{Casey13a}, which focused on successively widening symmetric search time windows around the GRB alerts considering a flat temporal signal probability density function.
% while the approach pursued here aims at identifying a systematically time shifted neutrino signal with respect to the electromagnetic emission.  
%The methods developed by the \IceCube collaboration searched for extended signals spread equally over each of the considered time windows, with the signal PDF described by a flat temporal component \cite{Casey13a}. 
In the case of a time-shifted signal, these methods suffer from reduced significance due to the accumulated background in the increasingly large time windows. 
In contrast, the technique presented here aims at identifying a time-shifted signal as a peak on top of flat background. 
\subsection{Potential physical delays considered}
\label{sec:physical_delays}
%The common time prof \mathcal{s} comprises a discrete representation of the stacked deviations of detection times $(\tau = t_{\nu} -  t_{\rm GRB})$ between the (first) detected photon signal $t_{\rm GRB}$, and the time of a possibly associated neutrino candidate $t_{\nu}$.
%However, different physical processes could lead to time shifts between the electromagnetic and the neutrino signal. 
%Consequently, the signatures of these processes might manifest most evidently in other stacked profiles than that for the generic shift of detection times $\tau$.
For maximum generality we perform a test for a constant offset $(\tau= t_{\nu} -  t_{\rm GRB})$ between the (first) detected photon signal $t_{\rm GRB}$ and the time of a possibly associated neutrino candidate $t_{\nu}$, for maximum generality.  In the case of a constant shift $\tau_{em}$ of the emission times of the neutrino with respect to photons at the source, it  translates into observed time delays at Earth $\tau_{\rm obs}$ that depend on the cosmological redshift $z$ of the GRB as: 
	\begin{align}
		 %\de t_{\rm obs} / \de t_{\rm em} &= R(t_{\rm obs}) / R(t_{\rm em}) = 1 +z \\
		  \label{eq:tobs_tem_z}
%		\tau_{\rm obs} &= \tau_{z} \cdot (1 + z) \; .
% \\
 		\tau_{\rm em} &= \tau_{\rm obs} / (1 + z) \; .
	\end{align}  
	%with the scale factor $R$.
	To test for these intrinsic time shifts, the distribution of $\tau_z = \tau / (1 + z)$ will be investigated. 
	%This modified delay in case of an observed excess would give ?????!!!
Note that the redshift is only measured for approximately 10\% of all GRBs, significantly reducing the statistics of the stacked profile when omitting all GRBs without determined redshift. 

%{\bf JS: I would leave out these details of LIV?} %JS
%Were asked by Dominik....
Effects due to LIV (see e.g. \cite{Amelino09a}, \cite{Jacob07a} and \cite{Amelino15a})  can also yield different arrival times at Earth for photons and neutrinos of high energy produced by a GRB. 
In a variety of quantum spacetime models, the velocity dispersion relation linking the energy of the particle $E$ and its momentum $p$ is modified by an additional term proportional to an integer power of the ratio of the energy to the Planck scale:
\begin{align}
	E^2-p^2c^2 &= \pm E^2 \cdot ( E/M_{\rm LIV} )^n \; ,
	\end{align}
	where $M_{\rm LIV}$ is the scale at which the symmetry is broken. The mass term $m^2 c^4$ can be neglected for neutrinos \cite{Jacob07a}. First-order terms with $n$=1 will be considered here as these exhibit the most sizeable effects.
% simply because you always start with first order? I would get rid of the details here! %JS 
%Had to be a bit more elaborated indeed. Once again this was asked by Dominik 
%, while  describing a wide range of models allows sizable departure from the Lorentz invariant case and hence detectable effects. 
Within this framework, the time shift observed at Earth will depend on the energy of the neutrino, the distance of the source $D(z)$ and the energy scale $M_{\rm LIV}$: 
	\begin{align}
	\Delta t_{\rm LIV} &= (\pm 1) \cdot E/M_{\rm LIV} \cdot D(z)/c \label{eq:liv} \; , 
	\end{align}
where $D(z)$ is the effective distance travelled by the particles taking into account the expansion of the Universe, and is defined according to \cite{Amelino15a} as:
\begin{align}
	D(z) &= \frac{c}{H_{0}}\int_{0}^{z} \frac{(1+z')dz'}{\sqrt{\Omega_m(1+z')^3+\Omega_\Lambda} } \; ,
 \end{align}
where $z$ is the redshift, $H_0$ is the Hubble constant, and $\Omega_m$ and $\Omega_\Lambda$ are the relative matter and dark energy densities of the Universe \cite{PDG}. %{\colr{red} Smthg about the 1+z factor in numerator}. 
These effects are expected to appear in a stacked histogram that accounts for both the estimated neutrino energy $E_{\rm est}$ and the distance of the source. 
Consequently, the variable to be probed is defined as:
	\begin{align}	
	\tau_{\rm LIV} &= \frac{\tau}{E_{\rm est} \cdot D(z)} \; , 
%		\propto \pm \frac{E}{E_{\rm est}} \cdot \frac{1}{M_{\rm LIV} \cdot c} \; .
	\end{align}
In case of a sizeable LIV effect, with a given value of $M_{\rm LIV}$ this yields
\begin{align}	
	\tau_{\rm LIV} &\propto \pm \frac{E}{E_{\rm est}} \cdot \frac{1}{M_{\rm LIV} \cdot c} \; ,
	\end{align}
and the time-stacked neutrino observations will accumulate around a single value of $\tau_{\rm LIV}$. In contrast, the distribution of events due to purely accidental coincidences will peak around zero. %since the range $\tau \in [-\taumax, +\taumax]$ is confined by the factor $1/E_{\rm est} \cdot D(z)$. 

The ratio $r= n_+ / n_-$ of spatially coincident events before and after the respective GRB alert is a very simple measure to probe the distributions while making the fewest assumptions on any model.
Any effect leading to different arrival times of neutrinos and gamma-rays from GRBs is expected to yield either positive or negative time shifts.
This ratio is calculated if both $n_+$ and $n_-$ are non-zero.%; for purely random coincidences, a mean value of $r=1$ is expected. 

Consequently, in the search for an associated neutrino signal from GRBs, three stacked time profiles for the measures $\tau$, \tauz and \tauliv were generated for all neutrino candidates which matched the coordinates of a reported GRB alert, and the ratio $r$ for the whole sample was computed.

\section{Implementation of the method and application to\Antares 2007-2012 data}
\label{sec:app_to_antares}
%
%%%%%%%%%% JULIA %%%%%%%%%%%
% 25.2.15
%
%\subsection{Analysis Parameters} \label{sec:time_stacking_parameters}
The expected number of background events $\mu_b$ increases with the solid angle of the search cones $\Omega(\deltamax)$ around each GRB's position and with the considered maximum time delay \taumax .
 %
%\begin{align}
%	\mub & \propto \Omega(\deltamax) \cdot  \taumax \; .
%	%\\ 	\mub & \propto \taumax
%\end{align} 
%
Hence, the choice of the search cone size and the probed time window should be optimised under reasonable physical considerations.% in order to stay as independent of any model assumption as possible and still allow even a faint signal correlated with electromagnetic GRB detections to protrude the background. 
%
%%%%%%%%%%%%%%%%%%%%
\subsection{Search Cone} \label{sec:coinc_cone}
%%%%%%%%%%%%%%%%%%%%
%
The determination of an optimally-sized search cone for spatially coincident neutrino candidates with a GRB alert was based on the maximisation of the ratio of signal to square root of background. 
Assuming a point-source-like signal at the GRB's location, the reconstructed neutrino directions approximately follow a two-dimensional Gaussian profile of standard deviation $\sigma_{\nu}$ around this position. This approach yields an optimum search cone size of $1.58 \cdot \sigma_{\nu}$ as derived for example in \cite{Alexandreas93a}. Note that the neutrino telescope resolution is usually stated as the median of the reconstructed direction error $m(\delta_{\rm err})$ (see for instance \cite{Antares12}). For angles in consideration here, the relation  $m(\delta_{\rm err})\sim 1.17 \sigma_{\nu}$ holds. 

The effects of uncertainty in GRB location $\mathrm{\Delta_{\rm err}}$ (sub-arc second for \SwiftUVOT or ground based telescopes, up to several degrees for \Fermi/GBM) is accounted for by extending the search window whenever $\mathrm{\Delta_{\rm err}} > \sigma_{\nu}$. 
%The effect of large uncertainties on the GRBs' positions, as given by the angular error boxes of the satellites $\Delta_{\rm err}$, should not be ignored, since these can vary from sub-arcseconds (from observations with the \SwiftUVOT instrument or ground-based telescopes) up to several tens of degrees for \FermiGBM alerts without any other follow-up observation.
%(in the worst case, for GRB110911, the \FermiLAT error box is as large as 50\deg). 
%Hence the size of the search window around each GRB is widened accordingly, whenever its respective error-box size $\Delta_{\rm err}$ exceeds the neutrino angular resolution $\sigma$.
%The search cone size is consequently defined as
%\begin{align}
%\delta_{\rm cut} &= 1.58 \cdot  \max (\sigma_{\nu}, \Delta_{\rm err}) 
%\label{eq:deltacut} \; . 
%\end{align}
As the contribution of random coincidences scales quadratically with $\Delta_{\rm err}$, the background in the cumulative profile might be dominated considerably by a few bursts with very large satellite error boxes. 
Consequently, a reasonable trade-off should be found. 
On the one hand, the statistics should not be reduced too much by excluding a large number of badly-localised bursts. 
On the other hand, the stacked timing profiles should not be dominated by one burst with a large error box, which naturally leads to many accidental spatial coincidences.
%Given, for instance, a search cone increased in size by a factor of 3 for a poorly localized burst, the associated coincident background would already dominate that of the bursts with small error boxes by one order of magnitude. 
In order to limit this effect without significantly reducing the data sample, a maximum search-cone size was chosen -- based on the distribution of $\rm{\Delta_{\rm err}^{\rm max}}$ shown in Figure 3 -- such that no GRB contributed more than an order of magnitude more of uncorrelated background than any other.
% one with the smallest associated search radii $1.58 \cdot \sigma_{\nu}$. 

The search-cone size is consequently defined as:
\begin{align}
\delta_{\rm cut} &= 1.58 \cdot  \max (\sigma_{\nu}, \ min (\rm{\Delta_{\rm err}, \Delta_{\rm err}^{\rm max} }) ) 
\label{eq:deltacut} \; . 
\end{align}
 Using the\Antares pointing resolution of $0.38\deg$, all bursts with error boxes larger than $\mathrm{\Delta_{\rm err}^{\rm max}}=1\deg$ were consequently  discarded from the search, which reduced the sample by $\sim 54\%$ while keeping $74\%$ of the total gamma-ray fluence of the sample, yielding  search-cone sizes in the range $[0.51\deg,1.58\deg]$. Note that Fermi-detected bursts with a resolution of $1^{\circ}$ are included.
\begin{figure} \centering
\includegraphics[width=0.495\textwidth]{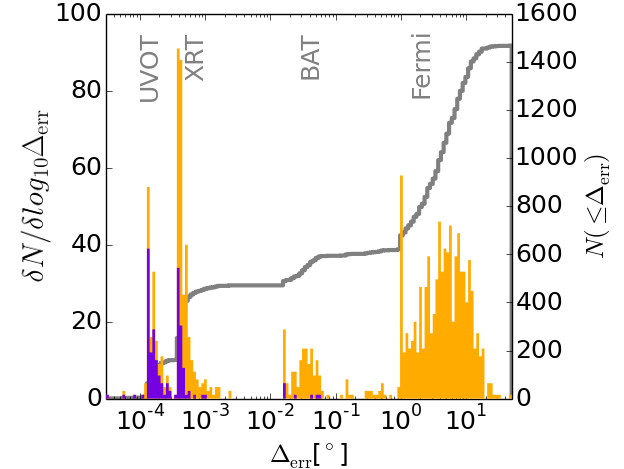}
\caption{\label{fig:err_radius}
%{\sl Upper panel}: 
Number of GRBs with a given error box $\mathrm{\Delta_{\rm err}}$ (orange). The cumulative distribution is shown by the grey line. For GRBs with measured redshift, the distribution is shown in violet.
%{\sl Lower panel}: Scatter plot of photon fluence $\mathcal{F}_\gamma$ of each GRB versus its error box $\Delta_{\rm err}$. 
%The black histogram shows the cumulative distribution of the fluence-weighted number of GRBs with an error box $\leq \Delta_{\rm err}$. 
%Gray dashed lines indicate the cut at $\Delta_{\rm err}$ as derived for the\Antares neutrino sample.
}
\end{figure}
%
%%%%%%%%%%%
\subsection{Maximum Time Delay} \label{sec:tau_max}
%%%%%%%%%%%
%The maximum considered time window should be limited to constrain the number of accidental coincidences from uncorrelated neutrino candidates in the GRB search cones.
The approach presented in this paper aims at being as model independent as possible. The maximum time shift anticipated from the astrophysical processes mentionned in section  \ref{sec:physical_delays} is used to set the time coincidence window. 
Intrinsic shifts in the emission times of neutrinos were predicted in \cite{Razzaque03a} with neutrinos $\sim 100 \s$ before the electromagnetic GRB signal. 
A precursor neutrino signal that might be emitted even tens of years before the actual GRB is derived in \cite{Granot03a}. 
Since the latter time scale exceeds the operational times of the current neutrino telescopes, we will omit this scenario. 
%Since these time scales can potentially cover or exceed the operation time of neutrino telescopes, the probed coincidence time window should be set to the whole period covered by data. But this would significantly reduce the detectability of shorter time scale phenomena, we will omit this scenario here.
Early afterglow emission of neutrinos $\sim 10\unit{s}$ after the burst are predicted in \cite{Waxman00b} and \cite{Murase07a} and extended neutrino fluxes up to 1 day after the prompt emission are derived in \cite{Razzaque13b}. 
These intrinsic time shifts between neutrino and photon signals are still well within the time scopes that have already been probed, for example in \cite{IceCube12a},\cite{Casey13a} -- without positive result.

Differences in arrival times induced by LIV effects would depend not only on $M_{LIV}$, but also on the energy of the particles and the distance of the source. 
However, a maximum expected time shift of neutrinos and photons can be inferred from Equation~\ref{eq:liv} using the existing limit on the LIV energy scale. 
The most stringent limit within the theoretical framework used here has been set to  $M_{\LIV} = 7.6 \cdot M_{\rm Planck}$ based on the \FermiLAT data \cite{Vasileiou13a}. 
%An equivalent but slightly weaker limit has been set by \cite{Vasileiou2015a} using a different approach.
%Neutrino telescopes are sensitive to neutrinos up to $\sim E_{\rm max}=10^9\GeV$. 
Using the distance $D(z)$ at a redshift of $z=8.5$, which is the highest measured redshift in the selected GRB catalog, and a maximum energy of $\sim E_{\rm max}=10^9\GeV$ accounting for the energy range at which a signal might be observed, a maximum time shift of $\taumax = 40$ days was derived. Even though the upper bound was derived from quantum space-time models, the search itself remains model independent.

A discretisation of the cumulative timing profiles into 150 bins was chosen, which allows time scales down to 13 hours to be probed. 
Given the low number of expected coincidences within the allowed time window (see section \ref{sec:time_stacking_pe}), this choice ensures that there will be much less coincidences than bins, leading to a quasi-unbinned approach \cite{Bose13a}. 
% YYY
\subsection{Final GRB Sample\label{sec:final_samples}}
Having chosen the maximal search time window and the largest angular search cone that should be taken into account, the final samples associated with the neutrino telescope data set were determined. 
The initially selected GRB catalog comprised 1488 bursts that had occurred between 2007 to 2012, which gives a detection rate of 0.68 bursts per day. 
Out of these, 563 have been selected for the search of associated neutrinos in the\Antares data using the criteria outlined above, of which 150 have a measured redshift $z$. 
%
%%%%%%%%%%%
%\section{Pseudo Experiments / Analysis Validations}
 %
%%%%%%%%%%%%%%%%%%%
\subsection{Statistical tests}
From the stacked histograms of $\tau$, $\tau_z$ and $\tau_{\rm LIV}$, test statistics are calculated that distinguish a systematically time-shifted neutrino signal associated with GRBs from the background-only hypothesis of purely accidental coincidences.
%The significance of an excess in the data is derived from comparing the measurement to the expectations from a large number of mere background realisations. 
A large number of background realisations preserving the telescope's acceptance are generated from the existing data sets by scrambling the time from the corresponding distribution of Fig. \ref{fig:event_distributions} and randomising the right ascension of detected neutrino candidate events in accordance with the flatness of the data distribution. 
%A large number of background realisations are generated from the existing data sets. 
%Neutrino candidates' right ascension is randomized which is justified by the flatness of the distribution shown in Figure~\ref{fig:event_distributions} and preserves the detector acceptance.  XXXXX
%The time of the events are scrambled, keeping the structure as presented in Figure~\ref{fig:event_distributions}.
The significance, of an excess in the data is then given by the $p$-value which is the probability to measure the test statistic in question (or more extreme values) from the background-only distribution.
%by a value $\Psi_{data}$ is then given by the probability to measure a value $\Psi\geq\Psi_{data}$ from the background-only distribution.

The test statistic associated to the ratio $r$ will be the variable itself. 
For the stacked histograms, 
%a more elaborated test statistic was proposed by \cite{Eijndhoven08a}: 
%the author introduced the 
the Bayesian observable $\psi$ to estimate the compatibility of a given stacked (and binned) time profile with the expectations from background has been proposed in \cite{Eijndhoven08a} and \cite{Bose13a}. 
This test statistic is proportional to the logarithm of the probability $p$ of an observation $D$ under an hypothesis $H$ %(in the case under study here that there is no signal) 
defined by a set of information $I$ (here that the stacked profile bins are filled following a multinomial law of known probabilities):
\begin{align} \label{eq:psi}
\psi 	&= - 10 \log_{10} p(D | H , I)  \nonumber \\
	   	&=-10 \left[   \log_{10} n! + \sum_{k=1}^{m} n_k \log_{10} p_k  - \log_{10} n_k!  \right] \; ,
			  %\; \text{for a uniform background hypothesis.} 
\end{align}
with $n$ events in the histogram in total, distributed in $k \in [1 \dots m]$ bins. 
The probability to fall within bin $k$ is $p_k$; for a uniform background distribution (i.e. in the case of the $\tau$ profile), $p_k = 1/m$ is simply given by the total number of bins $m$.% As ten times the logarithm of the ratio of two definite positive quantities, $\psi$ is usually expressed in dB \cite{Bose13a}.
For the non-uniform profiles \tauz and \tauliv, these probabilities have to be determined by a large number of pseudo-experiments simulating the background, of the order of $10^{7}$ to estimate the significance of a potential excess up to the $5\sigma$ level. 
% $\sim 4\cdot 10^6$
% from out_antares_40_159/pk/f0000/hist_tau.txt
The value of $\psi$ is calculated for each of the $\tau$,  $\tau_z$ and $\tau_{\rm LIV}$ profiles, correspondingly denoted $\psi$, \psiz and \psiliv.
%%Removed, suggestion of Marizio
%The binning of the histograms can, in principle, have a non-negligible influence on the distribution of the test statistic $\psi$ and thus the distinguishing power of a signal over background. 
%However, the number of coincident events in the stacked profiles is expected to be rather low, of the order of 10. 
%In this range, $\psi$ is discretely distributed since the histograms are filled with isolated events, and the choice of the bins' size has no considerable effect on the discovery power of the analysis. 
%We consequently chose a binning that isolates signals from different GRBs, while the emission associated with single bursts was basically comprised in one time bin. 

\label{sec:binning}
%
%
%Note that even if the presented technique is aimed at identifying the mere deviation of an observation from the hypothesis of randomized correlations instead of quantifying the observed time shift, the information of the most significant excess in the timing profile is still conserved and can be accessed in terms of the largest single contribution of a bin in Equation~\ref{eq:psi}. 
%Note that even if the presented technique is aimed at identifying a weak collective effect and not quantifying a potential shift, this information can still be retrieved.If a systematic shift is present it can be accessed in terms of the largest single contribution of a bin in Equation~\ref{eq:psi}.

%
%%%%%%%%%%%
%%%%%%%%%%%
%

\subsection{Sensitivity}
\label{sec:time_stacking_pe}
\label{sec:sensitivity}
Around $1.4\e{7}$ pseudo experiments yielding sky-maps of uncorrelated neutrino events were generated to simulate the case of purely accidental coincidences (background-only) between the\Antares neutrino data and the GRB catalogue. 
On average, 3.9 of the neutrino candidates are expected to match the bursts' search windows in time and space, with 0.7 of them coinciding accidentally with the bursts with measured redshift.

To illustrate the performance of the proposed technique to identify hypothetical neutrinos from GRBs,
% on a simple model with intrinsically time-shifted emission at the source, 
%a test signal was mimicked by associating neutrino candidates artificially with part of the GRBs at a hypothetical intrinsic time shift of $t_{\rm in}$ between 1 and 20 days.
a test signal was generated by associating neutrino candidates artificially with a fraction of the GRBs at a hypothetical intrinsic time shift of $t_{\rm in}$ with $t_{\rm in}=1,~5,~10~{\rm or~}20$ days.
That is, taking into account the cosmological redshift $z$, a simulated signal delayed by $t_\nu = t_{\rm GRB} + t_{\rm in} \cdot (1+z)$. 
%The size of this GRB sub-selection with a spatially coincident test-signal neutrino thus represented the signal strength. 
Its strength was quantified by the probability $f \in [0,1]$ that a GRB produced a signal in the neutrino telescope. The signal was consequently only simulated for those bursts for which the redshift could be determined, and the variable which has the best sensitivity to it will be $\tau_{\rm z}$ since signal will accumulate in one bin.

%%%%%%%%%%%
%\section{Signal Detection Power}
%%%%%%%%%%% 

%Modifs asked by DD and HC from here (simplification of sens. and limits) 
\begin{figure}[h] \centering
\includegraphics[width=0.495\textwidth]{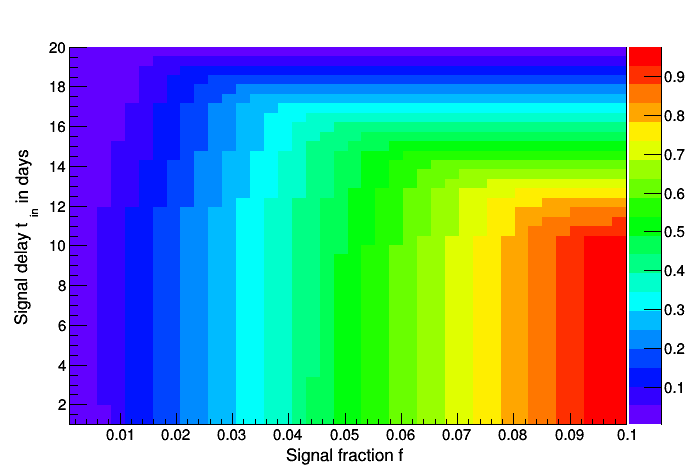}
\caption{\label{fig:eff_vs_time}
%{\sl Upper panel}: 
Detection efficiency (color scale) at the $3\sigma$ level using the \psiz test statistic as a function of the signal strength $f$ (see text) and intrinsic time delay at the source for a signal as described in section~ \ref{sec:time_stacking_pe}.
%{\sl Lower panel}: Scatter plot of photon fluence $\mathcal{F}_\gamma$ of each GRB versus its error box $\Delta_{\rm err}$. 
%The black histogram shows the cumulative distribution of the fluence-weighted number of GRBs with an error box $\leq \Delta_{\rm err}$. 
%Gray dashed lines indicate the cut at $\Delta_{\rm err}$ as derived for the\Antares neutrino sample.
}
\end{figure}

\begin{figure}[h] \centering
	\includegraphics[width=0.48\textwidth]{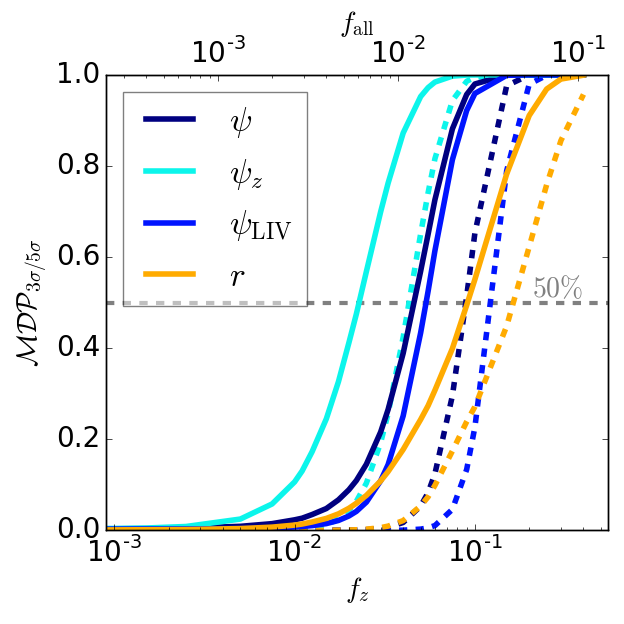}	
	\caption{\label{fig:results_antares_like_5sigma_2_1}	
	Detection probability \MDP at $3\sigma$ (solid) and $5\sigma$ (dashed lines) for the test statistics $\psi$, \psiz, \psiliv and $r$ as a function of the mean fraction $f$ of GRBs with one associated signal neutrino at $t_\nu = t_{\rm GRB} + 5 d \cdot(1+z)$.  
	The fraction $f_z$ denotes the fraction of GRBs in the sample with determined redshift $z$, whereas $f_{\rm all}$ gives the fraction of the whole sample.}
\end{figure}

\begin{figure}[h] \centering
	\includegraphics[width=0.48\textwidth]{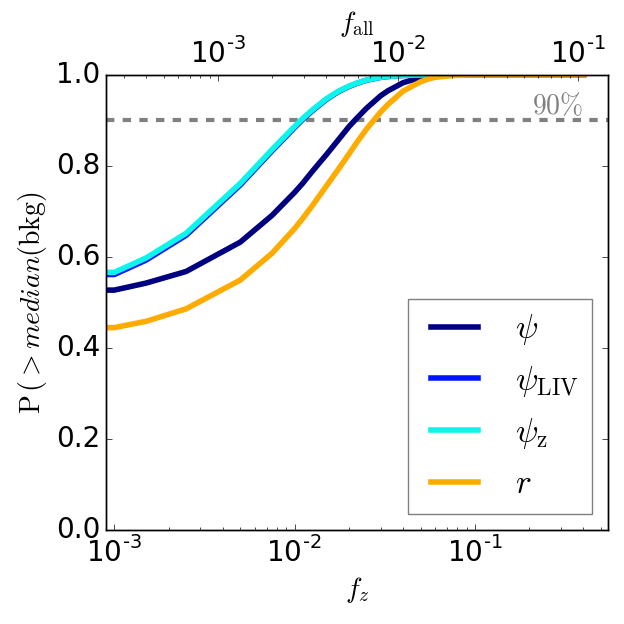}
	\caption{\label{fig:results_antares_like_5sigma_2_2}	
	Probabilities $P$ to measure values of the test statistics above the median value from the background-only realisations as a function of $f_z$  or $f_{\rm all}$ (as in Figure ~\ref{fig:results_antares_like_5sigma_2_1}). 
	The sensitivity is given by the signal fraction $f$ where the curves reach $90\%$ probability (grey dashed line).
	Note that the curves for $\Psiz$ and $\Psiliv$ lie on top of each other. 
Probabilities were derived using the\Antares data from 2007-2012.}
\end{figure}

%%%%%%%%%%%
The discovery probability \MDP at $n\sigma$ significance level for a given signal strength is given by the fraction of realisations that lead to values of the test statistics (here $r$, $\psi$, \psiz or \psiliv) above a threshold corresponding to a p-value at the $n\sigma$ level on the background-only realisations.
%\begin{align}
%$\MDP=P(Q > Q_{\rm thres} | background)
%\end{align}
It represents the efficiency of the analysis and the specific test statistic to identify a signal being associated with a fraction of GRBs.  
The detection efficiency of the \psiz test statistic is independent of the time shift of signal for delays up to 10 days, as can be seen in Figure~\ref{fig:eff_vs_time}.
% This can be understood from the fact that contributions of GRBs at redshift $z$ to the $\tau_z$ distribution will be in $[-40/(1+z),40/(1+z)]$ and that since $90\%$ of GRBs have a redshift below $z_{med}=3.5$, the background distribution of $\tau_z$ will be flat in $[-40/(1+z_{med}),40/(1+z_{med})]\sim[-9,9]$ which makes a potential signal protrude more easaly. Outside this range the distribution decreases non monotonically be cause the $1/(1+z)$ factor and the variation of the redshift distribution, making it less easy for a signal to protrude over background.
The evolution of the efficiencies for an example signal at an intrinsic time shift of 5 days as a function of the signal strength $f$ is shown in Figure~\ref{fig:results_antares_like_5sigma_2_1}, and is hence representative of shifts from 0 to 10 days. Signal strength corresponding to discovery probabilities are summarized in Table \ref{tab:sensitivities}, for the whole sample and for GRBs with measured redshift.
%Since the signal was generated for a mean fraction $f$ of the GRBs with measured redshift $z$, the fraction was corrected for the test statistics $\psi$ and ratio -- which are evaluated on the whole sample of GRBs --  by $N_{{\rm GRB}, z}/ N_{\rm GRB}$.
%The signal leading to discoveries at the $5\sigma$ level (dashed lines) must naturally be stronger than for $3\sigma$ (solid lines). 
For instance, using the $\psi$ test statistics, if only $f=1.3\%$ of the GRBs would give rise to an associated signal neutrino, it would produce an excess of $3\sigma$ significance with $50\%$ probability, whereas a stronger signal in $2.4\%$ of the bursts would be identified at the $5\sigma$ level. For the sample of GRBs with measured redshift, the \psiz test statistic only needs a fraction $f_z=4.5\%$ which is half of the signal fraction necessary with the $\psi$ test statistic for the same detection efficiency. 
%The measure $\psi$ as evaluated from the timing profile of the generic time delay $\tau$ identifies a signal very efficiently, both at $3\sigma$ and $5\sigma$ significance levels. 
%For instance, if only $1.3\%$ of the GRBs would give rise to an associated signal neutrino\footnote{That is $3.75\%$ of all GRBs with measured redshift $z$.}, 
%%$3\sigma$ evidence for this signal would already be found in $50\%$ of all measurements, 
%it would produce an excess of $3\sigma$ significance with $50\%$ probability, 
%whereas a stronger signal in $2.4\%$ of the bursts would be identified at the $5\sigma$ level (see gray dashed line marking the $50\%$ probability in Figure~\ref{fig:results_antares_like_5sigma_2}). 
%Being evaluated on the sample of GRBs with given redshift $z$ only, the measure \psiz is naturally more suited to identify the test signal that was simulated for these GRBs since the signal will accumulate in one bin. 
%In 50\% of the cases, the signal can be distinguished from the background if the delay occurs in 2.2\% of the GRBs with $z$ measured at $3\sigma$, and with $5\sigma$ in 4\% of the GRBs. 
%%

The introduced time-stacking technique is consequently capable of robustly finding at the $3\sigma$ level an intrinsically delayed neutrino emission from GRBs as long as it is associated with at least 3 of the 563 bursts. 
%The test statistic chosen to identify Lorentz invariance violation \Psiliv is, as anticipated, less powerful at identifying the simulated test signal, which was chosen to mimic an intrinsic delay of neutrino and photon emission at the source. 
%Figure~\ref{fig:results_antares_like_5sigma}, upper right panel, demonstrates clearly that the timing profiles do not change significantly with this type of test signal. % JS ??!!!
 
The probability of measuring values of the test statistics exceeding the median background value for different signal strengths is shown in Figure~\ref{fig:results_antares_like_5sigma_2_2}.%, with the respective numbers given in Table~\ref{tab:time_stacking_limits}. 
 The sensitivity at $90\%$ ($99\%$) confidence-level (CL) is defined as the $90\%$ ($99\%$) CL upper limit that can be placed on the signal strength when observing the median background (see gray dashed line marking $90\%$). 
The sensitivities at $90\%$ and $99\%$ CL of the proposed analysis for the given test signal simulating neutrino emission delayed by five days at the source in a mean fraction of all bursts are summarised in Table~\ref{tab:sensitivities}. 
%Withis $m(f_{{\rm all}}^{90\% {\rm CL}})=0.6\%$. 
For instance, at $90\%$~CL, considering only the sub-sample of bursts with determined redshift and the test statistics \Psiz and \Psiliv, the method is sensitive to a signal in only $1.1\%$ of the bursts.

%Similarly, if the confidence level is set to $99\%$ we obtain $m(f_{{\rm all}}^{99\% {\rm CL}})=1.2\%$ for all bursts and $0.7\%$ for bursts with measured redshift (corresponding to $2.5\%$ of the entire sample).
%	

%
\begin{table*} \centering %\small
\begin{tabular}{l|rr|rr|rr|rr  }
\hline   \hline 
 Test Statistic               & \multicolumn{2}{c |}{Sensitivity at 90\% CL} & \multicolumn{2}{c |}{Sensitivity at 99\% CL} & \multicolumn{2}{c |}{\MDP $3\sigma$} & \multicolumn{2}{c }{\MDP $5\sigma$ }\\
	& $f_{\rm all}$ & $f_z$ & $f_{\rm all}$ & $f_z$ & $f_{\rm all}$ & $f_z$ & $f_{\rm all}$ & $f_z$ \\
\hline 
 $r$   & 0.8\% &3.0\%	& 1.5\% & 5.5\%	& 2.4\% & 9.0\% 		& 4.5\%& 17\%	\\ 
 \hline
 $\psi$   & 0.6\% & 2.2\%	& 1.3\% & 5.0\%	& 1.3\% & 5.0\%	& 2.4\% & 9.0\%	\\ 
 \hline
 \psiz   & 0.3\% & 1.1\% & 0.8\% & 3.0\% &       0.6\% & 2.3\%       & 1.2\% & 4.5\%	\\ 
 \hline
 \psiliv   & 0.3\% & 1.1\%	& 0.8\% & 3.0\% & 1.5\% & 5.5\%	& 3.0\% & 12.5\%	\\ 
\hline  
\end{tabular}
\caption{\label{tab:sensitivities}
Sensitivities at 90\% and 99\% confidence level and detection probabilities at $3\sigma$ and $5\sigma$ with 50\% statistical power  for a signal delayed by 5 days at the source (see text) for the different test statistics expressed in terms of the fraction $f_{\rm all}$  of the GRB sample with detectable signal and the fraction $f_z$ of the GRB sample with measured redshift $f_z$.  
 }
\end{table*}

\begin{table*} \centering %\small
\begin{tabular}{ l| rrr |rr | rr |rr }
\hline   \hline 
$\nu$ data sample & $\tau_{\rm tot}$  & $N_{\rm events}$ & $m(\delta)$ & $\deltamax$& \taumax   &  $N_{\rm  GRB}$ & $N_{{\rm GRB} ,z} $ &   $n_{\rm coinc}$ & $n_{{\rm coinc}, z}$\\
&	[d] & & [\deg] & [\deg]& [d] &  & & \multicolumn{2}{c }{(uncorrelated)}\\ 
\hline  
\ANTARES (07-12)	& 2154	& 5516 & 0.38 & 0.51 -- 1.59 & 40 & 563 & 150 & 3.9 & 0.7 \\
IC40 (08-09)				& 408 		&  12876	&  0.70 & 0.95 -- 2.99 & 40& 60 & 12 &35.0 & 4.0  \\
\hline
\end{tabular}
\caption{\label{tab:time_stacking_statistics}
Total live-time of the considered neutrino telescope data sets $\tau_{\rm tot}$, respective number of neutrino candidate events $N_{\rm events}$ and respective median angular resolution $m(\delta)$. %The range of search cone sizes $\deltamax$ around each GRB is determined using Equation~\ref{eq:deltacut}, while the maximal search time window \taumax is fixed at 40 days. 
Samples of $N_{\rm GRB}$ GRBs are identified (out of which $N_{{\rm GRB} ,z} $ have measured redshifts) for the search of correlations. 
Assuming totally uncorrelated neutrino data, the mean numbers of coincident events that would be expected within the GRB's search windows $n_{\rm coinc}$ are also given. }
\end{table*}

\section{Results and Discussion}
The data collected by the\Antares telescope from the years 2007 to 2012 were analysed to search for neutrinos within the predefined angular and timing search windows associated with the GRB catalogue. 
None of the neutrino candidates in the data matched these search windows, where 3.9 would have been expected from background (0.7 coincidences were expected for the GRBs with measured redshift $z$). 
The measured values of the test statistics are thus zero, and the ratio $r=n_+/n_-$ is undefined.
The probability to observe no events coinciding with all GRBs is relatively small, with $P(0 | 3.9)=1.2\%$ (and $51.4\%$ for GRBs with measured $z$). 

We verified the under-fluctuation to be of statistical origin instead of intrinsic systematic effects in  the search methodology or the software. 
In particular, we derived the number of coincidences when increasing independently $\tau_{max}$ and $\delta_{max}$. Using these enlarged coincidence windows, the number of coincident data events is close to the expected number of coincidences from randomized data. 

\begin{table*} \centering

\hspace*{-0.9cm}
\begin{tabular}{ r|rr |r|rrrr } 
\hline  \hline 
$f_{\rm all}$ &   $P(>\psimeas)$  & $P(> \! m(\psi))$ & $f_z$ &  $P(> \! \psizmeas)$ &  $P(> \! m(\psiz))$ & $P(> \! \psilivmeas)$  & $P(> \! m(\psiliv))$\\
&$\psimeas=0$ & $m(\psi)=73.3$ & (all $z$)& $\psizmeas=0$ & $m(\psiz)=0$  & $\psilivmeas=0$ & $m(\psiliv)=0$\\ \hline
     0.0\% &       98\% &       50\% &      0.0\% &       48.5\% &       48.5\% &       48.2\% &       48.2\% \\ \hline 
     0.04\% &       99\% &       54\% &      0.15\% &       59\% &       59\% &       59\% &       59\% \\
     0.29\% &       99\% &       75\% &      1.1\% &       90\% &       90\% &       90\% &       90\% \\
     0.60\% &      100\% &       90\% &      2.3\% &       98\% &       98\% &       98\% &       98\% \\
     0.69\% &      100\% &       93\% &    2.6\% &         99\% &       99\% &       99\% &       99\% \\
     1.07\% &      100\% &       98\% &    4.0\% &         100\% &      100\% &     100\% &     100\% \\
     1.33\% &      100\% &       99\% &    5.0\% &         100\% &      100\% &     100\% &     100\% \\
     2.10\% &      100\% &      100\% &    8.0\% &         100\% &      100\% &     100\% &     100\% \\ 
\hline 
\end{tabular}
\caption{\label{tab:time_stacking_limits}Probabilities $P$ to yield values of the test statistic $Q \in [\psi, \Psiz, \Psiliv]$ above the measurement $Q_{\rm meas}$ and above the median value $m(Q)$ as expected from pure background realisations for different fractions $f_{\rm all}$ ($f_z$) of all GRBs (with measured redshift $z$) with one associated signal neutrino intrinsically shifter by 5 days at the source.}
\end{table*}

In Table~\ref{tab:time_stacking_limits}, the probabilities $P$ to measure test statistics above the measurements and the expected values from the median background realisations are given. This results in $99\%$ CL limits of $f_{all}=0.04\%$ and $f_z=2.6\%$, and a $90\%$ CL upper limit on $f_z$ of $1.1\%$. With the aforementioned under-fluctuation, the setting of a $90\%$ CL limit on $f_{\rm all}$ defined according to section \ref{sec:sensitivity} is not possible. A conservative option would be to set the limit equal to the sensitivity as in \cite{IceCube13b}.
Since this method does not make use of the information contained in the actual nonobservation, the resulting $90\%$ CL of $f_{all}=0.6\%$ is weaker than the standard $99\%$  CL of $0.04\%$, so the value of $0.04\%$ should be used for both $90\%$ and $99\%$ CL.

%%% replaced by the above
%In Table~\ref{tab:time_stacking_limits}, the probabilities $P$ to measure test statistics  above the measurements and the expected values from the median background realisations are given. 
%\footnote{Note that since $23\%$ of all realisations yield a ratio of exactly $1$, the fraction above (and below) the median value is $38.3\%$ instead of $50\%$}.
We can state a sensitivity of $m(f_{\rm all}^{90\% {\rm CL}})= 0.6\%$ of all GRBs ($2.2\%$ for those with measured $z$), which is the median upper limit on the fraction of bursts that contain a signal of the form $\tau_{\rm s}=5\unit{d} \cdot (1+z)$. % (using $m(\psi) = 73.3\unit{dB}$). 
Furthermore, we see that $99\%$ of all realisations with a signal fraction $f_{{\rm all}} = 0.04\%$ would yield higher $\psi$ than observed, so we can exclude such a signal with $99\%$ confidence. 
Regarding the sample of bursts with measured redshift $z$, the observation of zero events matched the median expectation from background, so we could exclude that $1.1\%$ of them produced a signal neutrino with a delay shape $\tau_{\rm s}=5\unit{d} \cdot (1+z)$ with $90\%$ confidence, in accordance with the sensitivity that had previously been derived.

%In conclusion, the test statistics \psiz and \psiliv, being calculated from the sample of GRBs with measured redshift $z$, lead to better sensitivity than considering the full sample. 

%This in turn allows, after non-observation of any excess, to exclude the associated signal neutrinos with $1\%$ of these bursts. 
%Yet since a considerable under-fluctuation is observed, the limit that can be derived from the sample of all GRBs is even more stringent ($f_{\rm all}<0.04\%$ with $99\%$ confidence) than the sensitivity that was expected, $m(f_{\rm all}^{90\% {\rm CL}})=0.6\%$.
%Yet since an under-fluctuation is observed, a limit can be derived from the sample of all GRBs of $f_{\rm all}<0.04\%$ with $99\%$ confidence.%, even more stringent than the sensitivity that was expected, $m(f_{\rm all}^{90\% {\rm CL}})=0.6\%$.

%%%% results IC40 %%%%%%%
\subsection{Application to the IceCube IC40 Data Sample}
The same parameter optimisation and search has been performed with the public data sample\footnote{\IceCube IC40 neutrino candidates are available at \url{http://icecube.wisc.edu/science/data/ic40/}} from an analysis searching for neutrino pointlike sources  \cite{IceCube11c} of the \IceCube observatory in its 40-string configuration. These data cover April 2008 to May 2009 and comprise 12877 neutrino candidates. %During this period, 60 GRBs have been singled out with only 12 redshift measurements.
The selection procedure of neutrinos and GRBs is the same as in \ref{sec:app_to_antares}. With a resolution of $0.7\deg$ \cite{Karle10a} it leads to 60 GRBs ( respectively 12 with measured $z$) 35 of which are expected to be in coincidence with neutrinos (respectively 4). 
%Considering an angular resolution of $0.7\deg$ \cite{Karle10a}, which leads to a maximum search-cone size of $2.99\deg$ according to Equation \ref{eq:deltacut}, 35 of the neutrino candidates are expected to coincide with all GRBs' search windows (and 4.0 with the GRBs that have redshift determinations). 
The different parameters summarising the \ANTARES and \IceCube samples, including the number of coincident events $n_{\rm coinc}$ that would be expected if the neutrino data was completely uncorrelated with the chosen GRBs (i.e., the background-only hypothesis) are given in Table \ref{tab:time_stacking_statistics}.\\
%
%
%\begin{figure*}\centering
%	\includegraphics[width=0.99\textwidth]{./ic40_unblind_tau_histograms_psi}
%	\caption{\label{fig:ic40_unblind}Number of neutrino candidates from the \IceCube IC40 data-taking period that coincided spatially with one of the gamma-ray-burst alerts as reported by the \Swift and \Fermi satellites and the GCN network with the relative time delays $\tau, \tauz$ and \tauliv  {\sl  \xspace (from left to right)}.} 
%	%Distribution of the relative time delays $\tau, \tauz$ and \tauliv between neutrino candidates from the \IceCube IC40 data-taking period and gamma-ray-burst alerts as reported by \Swift, \Fermi and the GCN network with which they spatially coincided.}
%	% from unblind_ic40.py
%\end{figure*}
%
%
The \IceCube GRB sample shows significantly lower statistics, due to the fact that the published data spans only around one year compared to almost six years in the \Antares sample. 
In addition, because of the location of the detectors on Earth, $87\%$ of the sky is visible for the \Antares detector with unequal coverage, whilst the \IceCube experiment covers the northern sky but at all times. 
It is also worth noting that, due to the larger instrumented volume of the detector, the \IceCube data set contains more neutrino candidates than the \Antares one, while covering a smaller time period in which less GRB alerts were recorded. Both samples therefore explore different statistical regimes. 
In the end, 42 of the neutrino candidates fall within the search windows, with 8 for the bursts with measured $z$.
%The timing profiles of these candidates are shown in Figure~\ref{fig:ic40_unblind} for the three investigated time measures $\tau, \tauz$ and \tauliv. 
This is a slight fluctuation above the expectations from background, with $p$-values of 13.5\% (whole sample) and 5.1\% (GRBs with measured redshift), yielding excesses of moderate 1.5$\sigma$ and 1.9$\sigma$ significances, respectively. 
The observation is compatible with no correlation of the \IceCube data with the chosen GRB sample. 
Moreover, the timing profiles show no indication for any preferred time delay. 
%Giving that exactly the same optimisation procedure and analysis software have been used for both neutrino telescope data samples, this non-significant excess points out that the under fluctuation seen with\Antares data is not coming from the method or the software.
The measured and expected values as well as the corresponding significance of the different test statistics for the two studied samples, are summarised in Table~\ref{tab:results}.

\begin{table*}
\centering
\hspace*{-1.2cm}
\begin{tabular}{l |rr|rrr|rrr|rrrr} \hline \hline 
& \multicolumn{5}{c |}{ANTARES 07-12} &  \multicolumn{7}{c }{\IceCube IC40 08-09}  \\
 & \multicolumn{2}{c |}{all GRBs} &\multicolumn{3}{c |}{GRBs w/ $z$} & \multicolumn{3}{ c |}{all GRBs} &\multicolumn{3}{c}{GRBs w/ $z$} \\\hline
 & $n_{\rm coinc}$  &  $\psi$ & $n_{\rm coinc}$  &\psiz & \psiliv  & $n_{\rm coinc}$ & $r$  & $\psi$ & $n_{\rm coinc}$ & \psiz & \psiliv \\
& &(dB)  &  &    & (dB) & (dB)  & & (dB) & & & (dB) & (dB) \\ \hline
Bgd mean & 4.4 & 77.4 & 0.7 & 11.9 & 4.5  &  1.1 & 35.0 & 371.3& 4.0 & 56.6 & 10.4\\
Bgd median &  4 & 73.3 & 0 & 0 & 0 & 35& 1.0 & 371.8 & 4 &  & 56.3 & 7.9 \\
Measurement & 0  & 0 & 0 & 0 & 0 & 42 & 0.4 & 416.0 & 8 & 1.1  & 93.9 & 8.8  \\
$P(>{\rm meas.})$	& 98.8\% & 98.8\% & 48\% & 48.6\%  & 48.6\%
& 10.4\% & 0.4 & 14.0\% & 2.1\% &    & 6.1\% & 45.1\%   \\ 
$P(\geq {\rm meas.})$	& 100\% &100\% & 100\% & 100\%  & 100\%
& 13.5\% &   & 14.0\% & 5.1\% &   &  6.1\% & 45.1\%  
\\ \hline 
%sensitivity & $f =0.8\%$ &  $f =0.6\%$ &  $f =1.1\%$&  $f =1.1\%$\\ \hline
\end{tabular}
\caption{\label{tab:results} Mean and median values of the different test statistics used in this analysis as derived in the pseudo experiments of background and in the measurement using the neutrino candidates as selected in the\Antares data from 2007 to 2012 and \IceCube data from the IC40-period from April 2008 to May 2009. 
The number of data events coinciding spatially with the respective GRB samples $n_{\rm coinc}$ are also given.
The probabilities $P(>{\rm meas})$ and the $p$-value, $P(\geq {\rm meas})$ give the fraction of background-only pseudo experiments that yield test statistics above and at and above the measurement. Being 10 times the logarithm of two definite positive quantities, the $\psi$ type test statistics are usually expressed in dB \cite{Bose13a}. Note that since there are no coincidences in the \ANTARES data sample, $r$ is not defined
%\todo{they differ only in the case of discrete distributions or zero.}
}
\end{table*}

%\begin{table*}[h!] \footnotesize
%\begin{tabular}{ r|rr |r|rrrr } 
%\hline  \hline 
%$f_{\rm all}$ &   $P(>\psimeas)$  & $P(> \! m(\psi))$ & $f_z$ &  $P(> \! \psizmeas)$ &  $P(> \! m(\psiz))$ & $P(> \! \psilivmeas)$  & $P(> \! m(\psiliv))$\\
%&$\psimeas=0$ & $m(\psi)=73.3$ & (all $z$)& $\psizmeas=0$ & $m(\psiz)=0$  & $\psilivmeas=0$ & $m(\psiliv)=0$\\ \hline
%%$f_{\rm all}$ &   $p(\psi)$  & $P(>m(\psi))$ & $f_z$ &  $p(\psiz)$ &  $P(>m(\psiz))$ & $p$  & $P(>m(\psiliv))$\\
%%(all GRBs)&$\psimeas=0$ & $m(\psi)=73.3$ & (all $z$)& $\psizmeas=0$ & $m(\psiz)=0$  & $\psilivmeas=0$ & $m(\psiliv)=0$\\ \hline
%\input{efficiencies}
%\hline 
%\end{tabular}
%\end{table*}

\section{Conclusion}
A powerful method has been presented to identify a neutrino signal associated with GRBs if it is shifted in time with respect to the photon signal. 
The signal is distinguished from randomly distributed data as a cumulative effect in stacked timing profiles of spatially coincident neutrinos in the data from the \Antares and \IceCube neutrino telescopes. 
%The discrepancy between the signal and background-only measurements has been quantified in terms of a test statistic $\psi$, which is based on the cumulative profiles. 

Estimating the behaviour of the search for a large number of simulated measurements using randomised sky maps of the neutrino events, and comparing these with the actual neutrino telescope data, significances of the observations were derived. 
Using data from the \Antares neutrino telescope between the years 2007 and 2012, a deficit of spatially coincident neutrinos with the selected gamma-ray-burst catalogue was reported, with $98.8\%$ of the randomised data leading to more coincidences between the neutrino data and the GRBs. The application of the method to the public \IceCube data in its 40 line configuration gives results compatible with the expectation from background.

The presented approach could have identified an intrinsically time-shifted signal even if only of the order of one in a hundred GRBs would have given rise to a single associated neutrino in the \Antares data.
%No assumption on any model for neutrino emission had been made in this study, yet the average signal strength of 0.01 (corresponding to one signal neutrino per 100 GRBs) can still be compared to the models that had been considered in the search for simultaneous emission presented in previous studies.x
%The average signal strength of 0.01 (corresponding to one signal neutrino per 100 GRBs) can be compared to the models that had been considered in the search for simultaneous emission presented in previous studies. 
This is above the detectable neutrino signal predicted by the NeuCosmA model \cite{Huemmer10a} that is on average only of the order of $\sim 2\cdot 10^{-4}$ in the\Antares detector, and only the strongest individual burst yields a neutrino detection rate exceeding 0.01 \cite{Antares13b}. 
%However,  the future \KMNeT telescope is expected to detect a handfull of neutrinos from the prompt emission of GRBs following state of the art production models, so that the method presented here will be capable of distinguishing time-shifted neutrino emission from GRBs, if the neutrino flux would be of the same order of magnitude than that predicted by recent models like NeuCosmA. 

%Regarding the \IceCube reduced public dataset from the IC40 detector analysed, the observed neutrino events that coincide with the gamma-ray-burst search windows exceed the number that would have been expected from randomized data. 
%However, even in the most significant case where eight candidates, instead of four, were observed in the search windows of the GRBs with measured redshift $z$, the surplus is still compatible with the background expectation.
%Application of this search to the public IC40 dataset showed no significant excess.

In conclusion, novel analysis techniques have been developed that increase the sensitivity of existing neutrino searches from GRBs to models of delayed neutrino emission and Lorentz Invariance Violation. They allow extending the search for neutrinos from GRBs
with time displacements of up to 40 days. It confirms the absence of a significant neutrino signal being associated with GRBs that has so far been measured in the simultaneous time windows.

\bibliographystyle{spphys}       % APS-like style for physics

\section{acknowledgements}
The authors acknowledge the financial support of the funding agencies:
Centre National de la Recherche Scientifique (CNRS), Commissariat \`a
l'\'ener\-gie atomique et aux \'energies alternatives (CEA),
Commission Europ\'eenne (FEDER fund and Marie Curie Program),
Institut Universitaire de France (IUF), IdEx program and UnivEarthS
Labex program at Sorbonne Paris Cit\'e (ANR-10-LABX-0023 and
ANR-11-IDEX-0005-02), Labex OCEVU (ANR-11-LABX-0060) and the
A*MIDEX project (ANR-11-IDEX-0001-02),
R\'egion \^Ile-de-France (DIM-ACAV), R\'egion
Alsace (contrat CPER), R\'egion Provence-Alpes-C\^ote d'Azur,
D\'e\-par\-tement du Var and Ville de La
Seyne-sur-Mer, France; Bundesministerium f\"ur Bildung und Forschung
(BMBF), Germany; Istituto Nazionale di Fisica Nucleare (INFN), Italy;
Stichting voor Fundamenteel Onderzoek der Materie (FOM), Nederlandse
organisatie voor Wetenschappelijk Onderzoek (NWO), the Netherlands;
Council of the President of the Russian Federation for young
scientists and leading scientific schools supporting grants, Russia;
National Authority for Scientific Research (ANCS), Romania; 
Mi\-nis\-te\-rio de Econom\'{\i}a y Competitividad (MINECO):
Plan Estatal de Investigaci\'{o}n (refs. FPA2015-65150-C3-1-P, -2-P and
-3-P, (MINECO/FEDER)), Severo Ochoa Centre of Excellence and MultiDark
Consolider (MINECO), and Prometeo and Grisol\'{i}a programs (Generalitat
Valenciana), Spain; Agence de  l'Oriental and CNRST, Morocco. We also
 acknowledge the technical support of Ifremer, AIM and Foselev Marine
 for the sea operation and the CC-IN2P3 for the computing facilities.
%\end{acknowledgements}
%\input{my_bib}
%\bibliographystyle{aa}
%\bibliography{./mnemonic,./aa_abbrv,./bibfile}
%\bibliography{biblio_epjc}

%]  %\twocolumn[
\end{document}